\documentclass[12pt]{revtex4-1}
\usepackage{graphicx}
\usepackage{geometry}
\usepackage{bm}
\usepackage{amsmath}
\usepackage{subfigure}
\usepackage{amssymb}%
\usepackage{tabularx} 
\usepackage[mathlines]{lineno}
\usepackage{lineno}

\usepackage{etoolbox}
\AtBeginEnvironment{tabularx}{\nolinenumbers}
\AtEndEnvironment{tabularx}{\linenumbers}

\begin{document}


\title{Effects of the Direct Space Charge on Transverse Coupled-Bunch Instabilities}
\author{Chao Li}
\altaffiliation[]{li.chao@desy.de}
\affiliation{Deutsches Elektronen Synchrotron, Notkestrasse 85, 22607 Hamburg, Germany}
\date{\today}
\begin{abstract}

The ultralow transverse emittance of diffraction-limited electron storage rings leads to high beam densities, making direct space-charge effects increasingly relevant despite their relativistic suppression at high energy. In this work, we investigate the impact of direct space charge on transverse coupled-bunch instabilities, using PETRA-IV as a representative example. A Vlasov solver, multi-bunch particle tracking, and analytical calculations in normalized phase space are combined to study the coupled-bunch mode spectrum, the nonlinear incoherent tune spread, and the associated Landau damping. Particle-tracking simulations show that the space-charge-induced tune spread can provide sufficient Landau damping to stabilize higher-order head-tail modes. Landau stability diagrams constructed from the analytical tune spread of a frozen bi-Gaussian beam predict the stabilization of these higher-order modes, in agreement with the tracking simulations. These results demonstrate the important role of direct space charge in transverse coupled-bunch dynamics. The analysis framework developed here can also be extended to collective-instability studies in both electron and hadron accelerators.

\begin{description}
\item[PACS numbers] 41.75.-i, 29.27.Bd, 29.20.Ej 
\end{description}
\end{abstract}

\maketitle

\section{Introduction}\label{sec-I}
As electron storage rings advance toward the diffraction-limited regime, their natural transverse emittance is substantially reduced, resulting in significantly higher beam densities. Transverse coupled-bunch instabilities remain an important limitation on the achievable beam current~\cite{ng2006physics,PhysRevSTAB.1.044201,buffat2022experimental,antipov2025space,ohmi2025space}. Transverse wakefields, particularly those associated with the resistive-wall impedance, can drive both intrabunch and interbunch coherent motion, thereby leading to unstable beam motions. The spectrum of these collective modes is strongly influenced by chromaticity and by the internal head--tail structure of the bunches~\cite{chao1993physics,PhysRevAccelBeams.19.124401,PhysRevAccelBeams.24.024402}.

The space-charge effect is often considered negligible in conventional high-energy electron storage rings because of its strong relativistic suppression. However, this assumption may no longer hold as storage rings approach the diffraction-limited regime~\cite{antipov2025space}. The resulting density-dependent direct space-charge introduces additional complexity: it modifies the incoherent betatron frequencies and generates a tune spread, thereby altering the internal bunch dynamics and potentially providing Landau damping of unstable modes~\cite{mohl1974landau,metral2004stability,PhysRevSTAB.18.074401,hofmann2017space,kornilov2010head}. In general, damping mechanisms tend to maintain a compact, high-density beam core and thus preserve a strong space-charge effect, whereas instability-driven emittance growth reduces the beam density and consequently weakens it. A self-consistent treatment of the beam dynamics therefore requires the interplay among impedance-driven coherent motion, chromaticity, radiation damping, and nonlinear space-charge to be considered simultaneously.

In this work, we investigate the effect of direct space charge on the transverse coupled-bunch instability, using the PETRA-IV lattice parameters summarized in Tab.~\ref{tab:lattice} as a representative example~\cite{Schroerig5056,Agapov2024}. In this study, only the main RF cavity is included in the system.  The key question addressed is the Landau damping provided by the space-charge-induced incoherent tune spread. In Sec.~\ref{sec-II}, we introduce the physical models and assumptions used to describe the coupled-bunch modes, incoherent tune spread, and Landau stability diagram. Section~\ref{sec-III} presents the impedance model and the numerical methods employed in the simulations. The simulation and analysis results are presented in Sec.~\ref{sec-IV}, where the Landau stability diagram constructed from the analytical tune spread of a frozen bi-Gaussian beam is shown to provide predictions consistent with the tracking simulations. Finally, the main conclusions and further discussion are summarized in Sec.~\ref{sec-V}.

\begin{table}[!htp]
\caption{Lattice parameters of PETRA-IV W.O harmonic cavity.}
\centering
\begin{tabular}{|l|c|c|c|}
\hline
 & Symbol & Unit & PETRA-IV \\
\hline
Energy                & $E$                  & GeV  & 6                      \\ \hline
Circumference         & $C$                  & m    & 2304                   \\ \hline
Harmonic number       & $h$                  &      & 3840                   \\ \hline
Tunes                 & $\nu_x,\nu_y,\nu_s$  &      & 135.18, 86.27, 0.0048 \\ \hline
Momentum compaction   & $\alpha_c$           &      & $3.33\times10^{-5}$    \\ \hline
Synch. damping times  & $\tau_x,\tau_y,\tau_z$ & ms & 18, 22, 13           \\ \hline
Equilibrium emittance & $\epsilon_0$         & pm   & 20                     \\ \hline
Bunch length          & $\sigma_t$           & ps   & 8.0                    \\ \hline
\end{tabular}
\label{tab:lattice}
\end{table}

\section{Analysis formalism}\label{sec-II}
\subsection{Coupled-bunch instability and chromaticity effects}\label{sec-II-A}
Assuming that the ring is uniformly filled with equally spaced electron bunches having a prescribed longitudinal profile, the coherent frequency shift and growth rate of coupled-bunch modes can be estimated from the transverse impedance $Z_1^{\perp}$ as~\cite{chao1993physics}
\begin{equation}
\begin{split}
\label{eq:5.2}
\Omega^{\mu,l} - \omega_{\beta} - l \omega_s
&\approx
-i \frac{\Gamma(l+1/2)}{4\pi 2^l l!}
\frac{N r_0 c}{\gamma T_0 \omega_{\beta} \sigma_t}
\frac{\sum_{p} Z_1^{\perp}(\omega') h_l(\omega' - \omega_{\xi})}
{\sum_{p} h_l(\omega' - \omega_{\xi})}.
\end{split}
\end{equation}
Here, $\Gamma$ is the Gamma function, $r_0$ is the classical electron radius, $N$ is the number of electrons per bunch, $c$ is the speed of light, $\gamma$ is the Lorentz factor, $T_0$ is the revolution period, $\omega_0=2\pi f_0$ is the revolution angular frequency, $\omega_{\beta}=2\pi\nu_{u}f_0$ ($u=x,y$)  is the transverse betatron angular frequency, $l$ is the azimuthal mode index, $\omega_{\xi}=\xi \omega_{0}/\eta$ is the spectral shift due to the chromaticity and $\eta=\alpha_c-1/\gamma^2$  is the slip factor. The impedance is sampled at $\omega'=\omega_{\beta}+(pM+\mu)\omega_0$, where $p\in\mathbb{Z}$, $M$ is the number of equally spaced bunches, $\mu$ is the coupled-bunch mode index ranging from 0 to $M-1$, The impedance weighted by the mode spectrum in the last factor of Eq.~(\ref{eq:5.2}) is commonly referred to as the transverse effective impedance. For a Gaussian longitudinal beam profile, the power spectrum of the azimuthal mode $l$ is
\begin{equation}
\label{eq:5.1}
h_l=(\omega\sigma_t)^{2l}e^{-(\omega\sigma_t)^2}.
\end{equation}
For a rigid bunch, Eq.~(\ref{eq:5.2}) reduces to
\begin{equation}\label{eq:2.1}
\begin{split}
(\Omega^{\mu}-\omega_{\beta})_{\perp}
&=
-i \frac{MNr_0c}{2\gamma T_0^2\omega_{\beta}}
\sum_{p=-\infty}^{\infty} Z_1^{\perp}(\omega').
\end{split}
\end{equation}

\begin{figure}[!htp]
\centering
\includegraphics[width=0.24\textwidth]{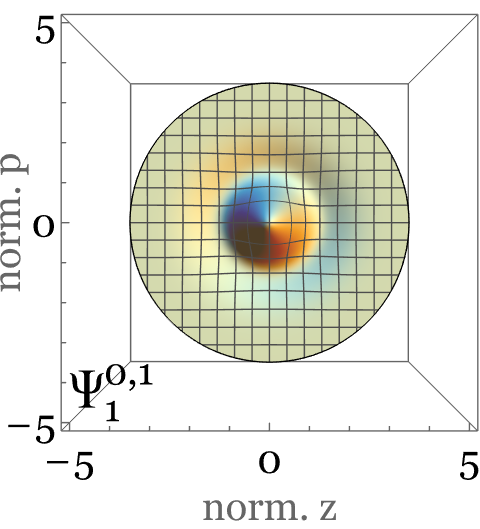}
\includegraphics[width=0.24\textwidth]{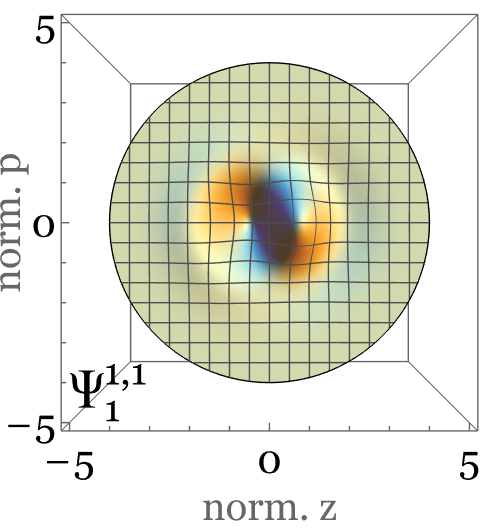}
\includegraphics[width=0.24\textwidth]{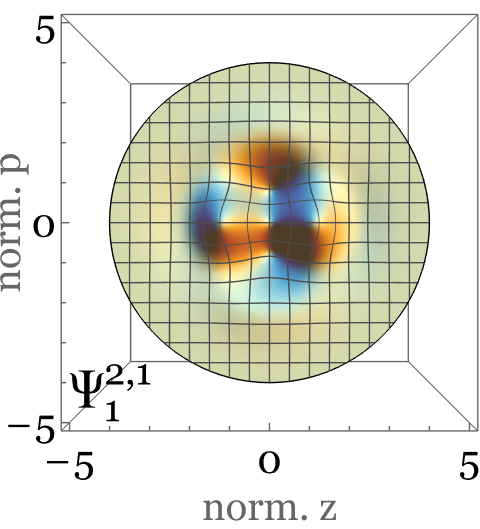}
\includegraphics[width=0.24\textwidth]{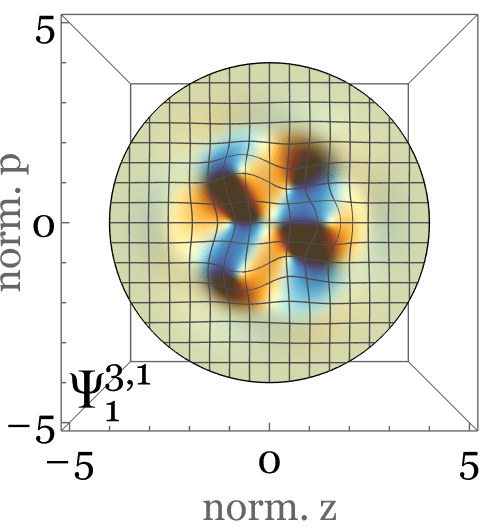}
\caption{Laguerre--Hermite mode structures in longitudinal phase space obtained from Eq.~(\ref{eq:5.4}). From left to right, the azimuthal mode index $l$ varies from 0 to 3, while the radial mode index is fixed at $k=1$.}
\label{fig:ana_mode_phase_structure}
\end{figure}

According to Eq.~(\ref{eq:5.2}), the instability growth rate is determined by the overlap between the transverse impedance and the mode spectrum. A nonzero chromaticity shifts the mode spectrum by $\omega_{\xi}$ and can therefore reduce the growth rate of unstable modes. For a machine with positive slip factor $\eta$, increasing the chromaticity from zero to positive values stabilizes the mode zero while higher-order azimuthal modes become dominant sequentially~\cite{chao1993physics}. However,
Eq.~(\ref{eq:5.2}) relies on several assumptions: (i) the bunch profile remains close to the unperturbed distribution, (ii) the beam current is sufficiently low that strong mode coupling does not occur, and (iii) radial modes are neglected. The first assumption limits the validity of the analytical mode decomposition when the bunch profile is strongly distorted during the evolution. The second makes Eq.~(\ref{eq:5.2}) a low-current approximation and prevents it from describing mode-coupling instabilities. The third assumption would make the results inaccurate when radial modes are significantly excited. A more general treatment requires solving the Fokker--Planck equation or a set of coupled Vlasov equations based on a phase-space expansion or discretization~\cite{Li:2022neg,berg1996coherent,PhysRevAccelBeams.24.024402,PhysRevAccelBeams.21.114404,PhysRevSTAB.17.021007,Antipov:2023xtx}. For a Gaussian longitudinal beam distribution, the Hermite--Laguerre basis can be used, as discussed in Eq.~(6.198) and Eq.~(6.238) in Ref.~\cite{chao1993physics}. The spectral component of mode $(l,k)$ is
\begin{equation}\label{eq:5.3}
\begin{split}
g_{lk}(\omega)
=
\frac{1}{\sqrt{2\pi k!(l+k)!}}
\left(\frac{\omega\sigma_t}{\sqrt{2}}\right)^{l+2k}
e^{-\omega^2\sigma_t^2/2},
\end{split}
\end{equation}
and the corresponding perturbation in longitudinal phase space is
\begin{equation}\label{eq:5.4}
\begin{split}
\Psi_1^{l,k}e^{-\Omega s/c}
\propto
e^{-r^2/2}r^lL_k^l(r^2/2)\cos(l\phi)e^{-\Omega s/c},
\end{split}
\end{equation}
where $r$ and $\phi$ are the normalized longitudinal phase-space coordinates, $k$ is the radial mode index, and $L_k^l$ is the associated Laguerre polynomial. Figure~\ref{fig:ana_mode_phase_structure} shows the corresponding mode structures for  $k=1$, $l\leq3$.

\subsection{The direct space-charge effect and tune spread}\label{sec-II-2}

The direct transverse space-charge force is intrinsically nonlinear and defocusing. It produces an amplitude-dependent incoherent tune depression and a corresponding tune spread across the beam. In addition, the transverse charge density varies along the bunch, introducing a longitudinal dependence of the tune shift. The resulting tune footprint modifies the resonance conditions and can increase the overlap with incoherent resonance lines. At the same time, the tune spread can provide Landau damping and thereby stabilize unstable collective modes.

If the frozen model is applied, the space-charge force can be represented as a nonlinear lens, allowing to get the incoherent tune shift and tune spread analytically. Assuming a three-dimensional Gaussian beam distribution, restricting the discussion to the transverse dynamics, following the treatment of Ref.~\cite{PhysRevAccelBeams.25.121001}, in Courant--Snyder action--angle frame, the Hamiltonian can be written as
\begin{equation}\label{eq:2-2-1}
H(J_x,J_y,\phi_x,\phi_y;s) =
\sum_{u=x,y}\nu_{u0}J_u +
U(J_x,J_y,\phi_x,\phi_y;s),
\end{equation}
where $U$ denotes the potential from the direct space-charge and $\nu_{u0}$ is the bare tune. Correspondingly, the normalized transverse bi-Gaussian distribution in action space is
\begin{equation}
\label{eq:4-3-1}
F(J_x,J_y)
=
\prod_{u=x,y}
\frac{1}{J_{u,\mathrm{rms}}}
\exp\left(-\frac{J_u}{J_{u,\mathrm{rms}}}\right),
\qquad
J_{u,\mathrm{rms}}=\epsilon_{u,\mathrm{eq}}.
\end{equation}

Introducing the azimuthal coordinate $\theta=2\pi s/C$ and expanding the space-charge potential in a Fourier series in $\phi_x$ and $\phi_y$ gives
\begin{equation}\label{eq:2-2-2}
U(J_x,J_y,\phi_x,\phi_y;\theta)
=
\sum_{m_x,m_y}
\widetilde{U}_{m_x,m_y}(J_x,J_y;\theta)
e^{i(m_x\phi_x+m_y\phi_y)},
\end{equation}
with Fourier coefficients
\begin{equation}\label{eq:2-2-3}
\widetilde{U}_{m_x,m_y}(J_x,J_y;\theta)
=
\frac{C}{2\pi}
\frac{\lambda r_0}{\gamma^3\beta^2}
\int_{0}^{\infty}\mathrm{d}q\,
\prod_{u=x,y}
\frac{
(-1)^{m_u/2}
e^{-w_u+i m_u(\chi_u-\nu_{u0}\theta)}
I_{m_u/2}(w_u)
}{
\sqrt{2\sigma_u^2+q}
},
\end{equation}
where $\chi_u$ is phase advance decided by the lattice, $\sigma_u$ is the rms beam size, $I_{m_u/2}$ is the modified Bessel function, $w_u=J_u\beta_u/(2\sigma_u^2+q)$, and $\beta_u$ is the betatron function. The phase-independent component $\widetilde{U}_{0,0}$ determines the amplitude-dependent incoherent tune shift,
\begin{equation}\label{eq:2-2-4}
\delta\nu_u(J_x,J_y)
=
\frac{1}{2\pi}
\oint
\frac{\partial\widetilde{U}_{0,0}(J_x,J_y)}
{\partial J_u}
\,d\theta.
\end{equation}
The phase-dependent Fourier components act as resonance-driving terms. The line density $\lambda$ in Eq.~(\ref{eq:2-2-3}) depends on the longitudinal position. For an equilibrium Gaussian longitudinal distribution, averaging the local space-charge tune shift over the longitudinal distribution introduces an additional reduction factor of $1/\sqrt{2}$ relative to the peak line-density value.

\subsection{Dispersion relationship and Landau damping}\label{sec-II-3}
The incoherent tune spread can provide Landau damping through phase mixing among particles with different betatron frequencies and may therefore suppress coherent beam oscillations. Following Ref.~\cite{Berg:1996ez}, the corresponding stability diagram can be written as
\begin{equation}
\label{eq:2-3-1}
1=-\Delta\nu_{u,\mathrm{coh}}
\iint\frac{J_u\,\partial F(J_x,J_y)/\partial J_u
}{\nu+\delta\nu_u(J_x,J_y)+i0^{+}}
\,dJ_x\,dJ_y,
\qquad u=x,y,
\end{equation}
where $i0^{+}$ denotes an infinitesimally small positive imaginary part defining the Landau contour. The stability boundary is obtained by scanning the real parameter $\nu$ and evaluating the corresponding complex coherent tune shift $\Delta\nu_{u,\mathrm{coh}}(\nu)$. The resulting trajectory in the complex $\Delta\nu_{u,\mathrm{coh}}$ plane defines the Landau stability diagram. With the sign convention adopted here, a coherent mode is expected to be Landau damped when its coherent tune shift lies inside the stable region bounded by the diagram.

It is worth noting that the Landau Stability diagram from Eq.~(\ref{eq:2-3-1}) is centered with respect to the beam center ($J_x=J_y=0$), while the incoherent tune is depressed by the space charge.  Therefore, a corresponding shift of the Landau stability diagram is required when the collective modes are expressed with respect to the bare tune. The shift of the  Landau Stability diagram can be estimated by 
\begin{equation}
\label{eq:2-3-2}
\Delta\nu_{u,s} = \iint F(J_x,J_y) \delta\nu_u(J_x,J_y) \,dJ_x\,dJ_y, \qquad u=x,y.
\end{equation}
Since direct space charge is defocusing, the Landau stability diagram is consequently shifted
as a whole toward lower tunes.

\section{Impedance model and space-charge solver}\label{sec-III}

\subsection{Impedance and wake model}\label{sec--III-A}
The resistive-wall (RW) impedance and wakefields of an infinitely thick metallic beam pipe of radius $b$ and conductivity $\sigma$ are well known; the corresponding expressions are given, for example, in Eqs.~(2.22, 2.53, 2.75) in Ref.~\cite{chao1993physics}. For an elliptical beam pipe, the Yokoya form factors~\cite{yokoya1993resistive} can be applied to account for the chamber ellipticity. Here we apply a simplified RW model as summarized in Tab.~\ref{tab:rw_para}~\cite{LI2024170031}, where ID represents the Insert Device. The total transverse RW impedance and wake are obtained by summing the contributions of the individual sections with weights determined by the local betatron functions. 

\begin{table}[!htp]
\centering
\caption{Simplified RW sections in PETRA-IV.}
\label{tab:rw_para}
\begin{tabular}{|l|c|c|c|c|c|c|}
\hline
Section & Number & Length / m & Gap / mm
& $\bar{\beta}_x$ / m
& $\bar{\beta}_y$ / m
& $\sigma$ / $\Omega^{-1}\,\mathrm{m}^{-1}$ \\
\hline
5 mm ID & 4  & 5    & 5  & 3.14 & 3.14 & $2.5\times10^{7}$ \\ \hline
6 mm ID & 17 & 5    & 6  & 3.14 & 3.14 & $2.5\times10^{7}$ \\ \hline
7 mm ID & 5  & 10   & 7  & 6.08 & 6.08 & $2.5\times10^{7}$ \\ \hline
Ring    & 1  & 2149 & 20 & 2.71 & 4.25 & $5.9\times10^{7}$ \\ \hline
\end{tabular}
\end{table}

\subsection{Space-charge solver benchmark}\label{sec-III-B}
In principle, a fully three-dimensional PIC solver with open boundary conditions provides the most general treatment of the direct space-charge fields. In electron storage rings, however, radiation damping and quantum excitation generally lead to an approximately bi-Gaussian transverse beam distribution, while the longitudinal motion is slow compared with the transverse betatron motion. Moreover, the longitudinal space-charge force scales as $1/(\gamma\sigma_z)^3$ and can therefore generally be neglected for the present parameters. In the beam rest frame, the bunch is highly elongated, such that $\gamma\sigma_z\gg\sigma_u$ ($u=x,y$). Under this condition, the longitudinal dependence of the transverse space-charge field can be factorized, as shown in the Appendix, and enters primarily through the local line density $\lambda(z)$~\cite{4440457}.

Accordingly, the 2.5D space-charge solvers implemented in Elegant~\cite{elegantWebPage} and CETASIM~\cite{LI2024170031}, based on the Bassetti--Erskine formulation, are employed for the numerical simulations. The bunch is divided into longitudinal slices, and the transverse space-charge kick in each slice is calculated independently using the corresponding local line density. This approach substantially reduces the computational cost while retaining the dominant transverse space-charge effects and enables a direct benchmark against the theoretical predictions.

\begin{figure}[!htp]
\centering
\includegraphics[width=1\textwidth]{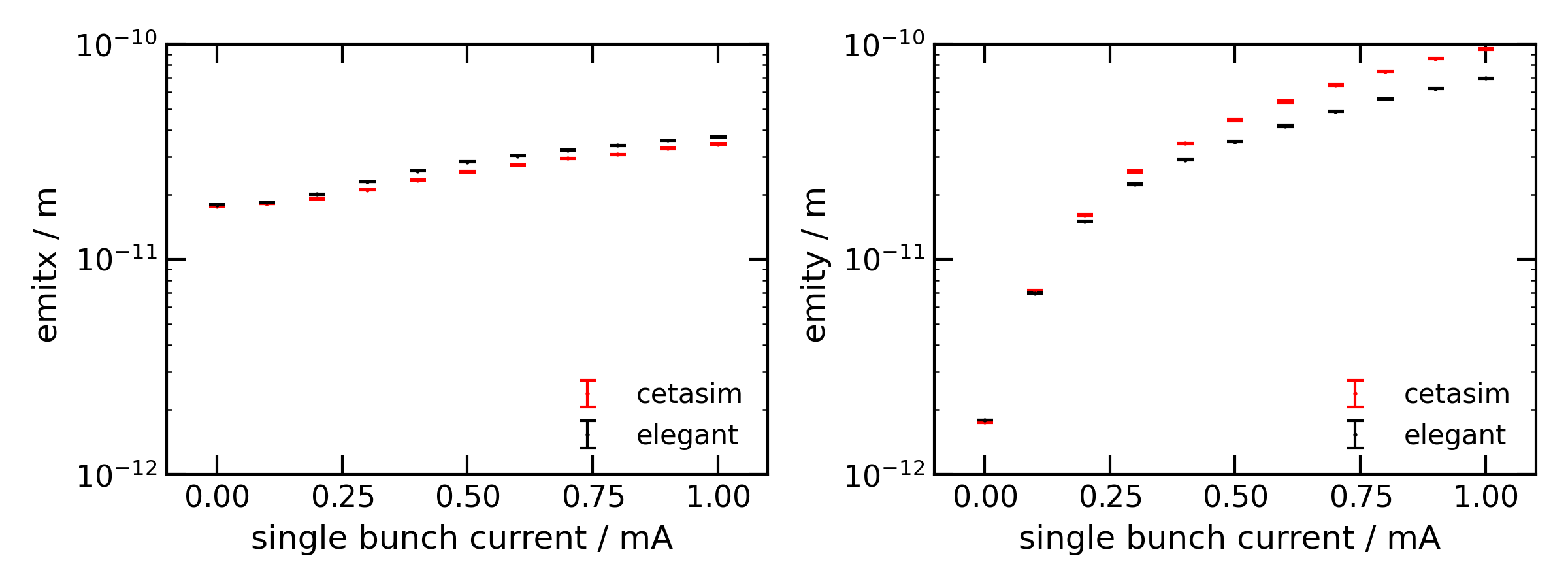}
\caption{Rms emittance as a function of single-bunch current obtained with Elegant and CETASIM using the one-turn-map tracking model. The emittance growth in both planes is exaggerated compared with element-by-element tracking.}
\label{fig:sc_emit_elegant_cetasim}
\end{figure}

Figure~\ref{fig:sc_emit_elegant_cetasim} compares the equilibrium horizontal and vertical emittances as a function of single-bunch current after 50,000 turns of tracking, with the ring represented by a linear one-turn-map. For both codes, the mean emittances are calculated over the final 1,000 turns, and the corresponding rms variations are shown as error bars. The two codes show good agreement in the overall dependence of the emittance on single-bunch current. Noticeable discrepancies in the vertical emittance appear as the current approaches 1~mA. These differences are mainly attributed to halo formation in transverse phase space, which can contribute significantly to the second-order moments. Care should therefore be taken when emittance growth or detailed incoherent resonance structures are used as figures of merit for assessing space-charge effects. In particular, structures observed with the one-turn-map model can be smeared out or suppressed in more realistic element-by-element tracking.

\subsection{Simplification in simulation}\label{sec-III-C}
For transverse coupled-bunch simulations, a large number of bunches and macroparticles per bunch must be tracked. A complete element-by-element beam-dynamics model incorporating short- and long-range collective effects together with space-charge forces is therefore computationally intensive. In this study, the short-range wakefields are represented by a broadband impedance, whereas the long-range wakefields are treated in the time domain, with the intrabunch variation of the long-range wake neglected. The broadband component of the RW impedance is discretized using $2^{16}$ frequency bins up to 500~GHz, while the long-range wakefield from the preceding 20 turns is retained. For all simulation configurations, per bunch includes 10,000 macroparticles.

To reduce the computational load further, we employ a simplified one-turn-map model in which the linear lattice is represented by a one-turn transfer matrix and the integrated space-charge effect by a single lumped kick per turn. The betatron function at the space-charge kick is chosen as $\beta_u=R/\nu_u$. Lattice nonlinearities can be incorporated through amplitude-dependent tune shifts, while chromatic effects are included to first order. This model retains the essential ingredients required to study coupled-bunch instabilities and substantially reduces the computational load. Nevertheless, element-by-element tracking remains necessary when space-charge-induced emittance growth or detailed incoherent resonance structures are used as figures of merit, since the local interplay between the lattice and space-charge forces cannot be fully captured by a single lumped kick. The one-turn-map model should therefore be regarded as a simplified and, in some cases, conservative representation of the space-charge dynamics.

\section{Simulations and analysis}\label{sec-IV}
\subsection{Coupled-bunch modes without space charge}\label{sec-IV-A}
\begin{figure}[!htp]
\centering
\includegraphics[width=1\textwidth]{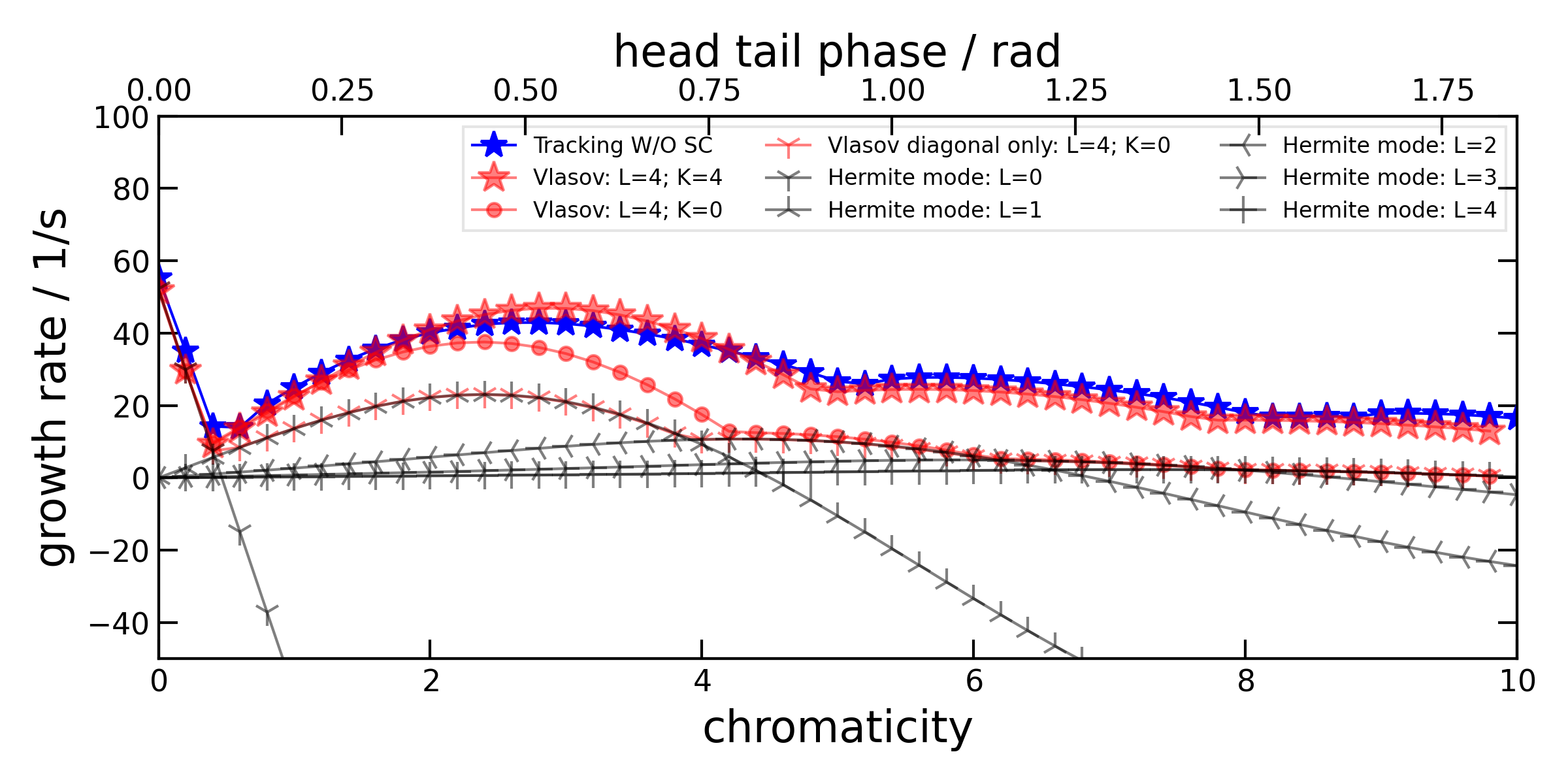}
\caption{Benchmark of the instability growth rates obtained using different analytical methods. The blue curve represents particle-tracking results, the red curves show predictions from the Vlasov solver, and the black curves correspond to Eq.~(\ref{eq:5.2}). The ring is uniformly filled with 80 bunches at a total beam current of 10~mA.}
\label{fig:growth_rate_benchmark}
\end{figure}

As explained in Sec.~\ref{sec-II-A}, under the assumption of a Gaussian beam distribution, chromaticity introduces a spectral shift that can reduce the instability growth rate.  Fig.~\ref{fig:growth_rate_benchmark} compares the growth rates obtained from tracking simulations with those predicted by different analytical approaches. In the tracking simulations, the ring is uniformly filled with 80 bunches at a total beam current of 10~mA. Synchrotron-radiation (SR) damping and quantum excitation are switched off, while the RF voltage and phase are adjusted to maintain the nominal synchrotron tune. The black curves show the growth rates of different azimuthal modes $l$ obtained from Eq.~(\ref{eq:5.2}). For comparison, the Vlasov solver is applied to study different levels of mode coupling. When only azimuthal modes are retained and all off-diagonal elements of the interaction matrix are artificially set to zero (Vlasov, diagonal only; $l=4$, $k=0$), the resulting growth rates coincide with those obtained from Eq.~(\ref{eq:5.2}). Retaining the off-diagonal terms (Vlasov; $l=4$, $k=0$) includes coupling among the azimuthal modes and, in particular, increases the growth rate of the $l=1$ mode. When radial modes are also included (Vlasov; $l=4$, $k=4$), substantially better agreement with the tracking results is obtained. For the parameter range considered here, $l=4$ and $k=4$ are sufficient to ensure good convergence of the Vlasov calculation.

\begin{figure}[!htp]
\centering
\includegraphics[width=0.48\textwidth]{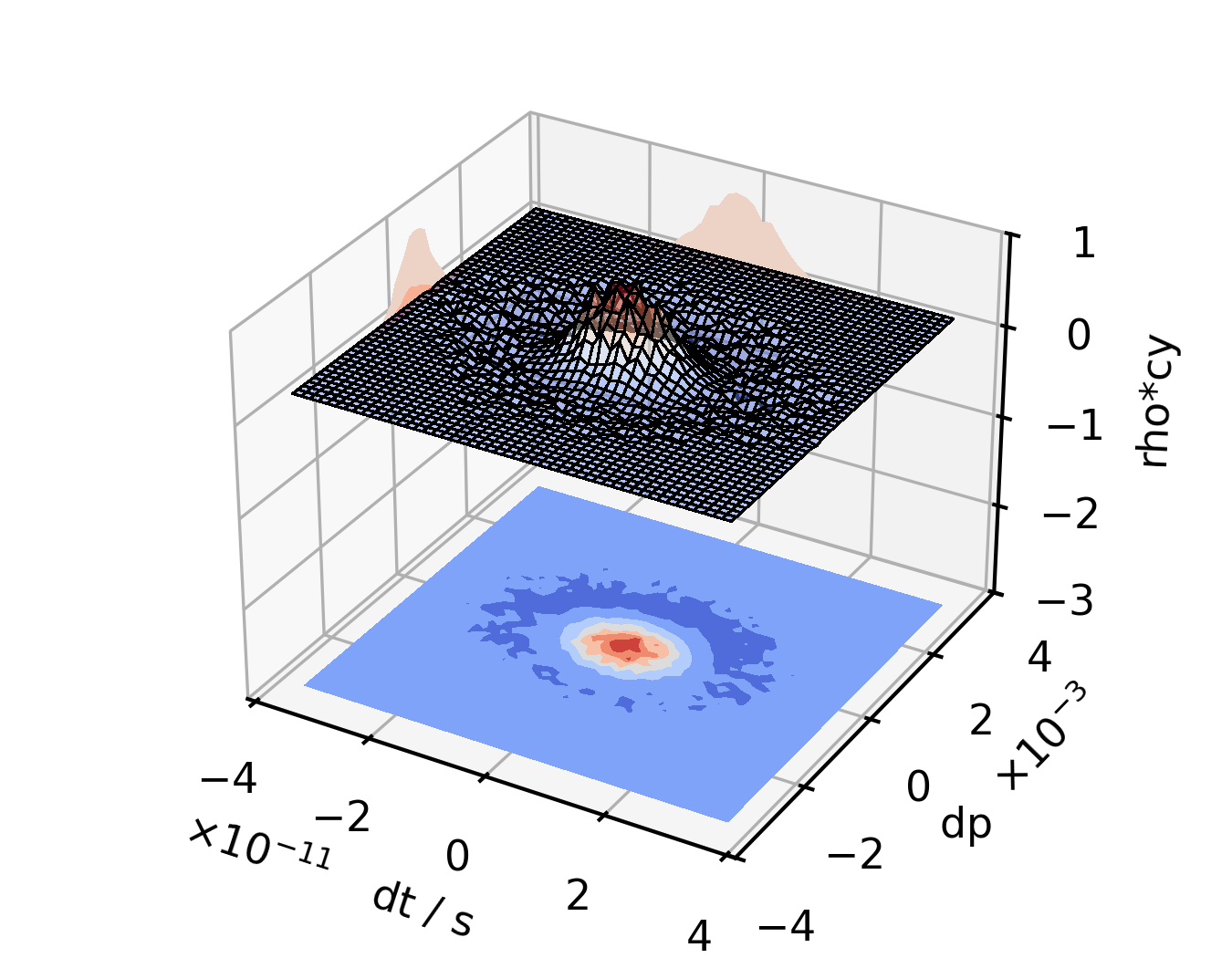}
\includegraphics[width=0.48\textwidth]{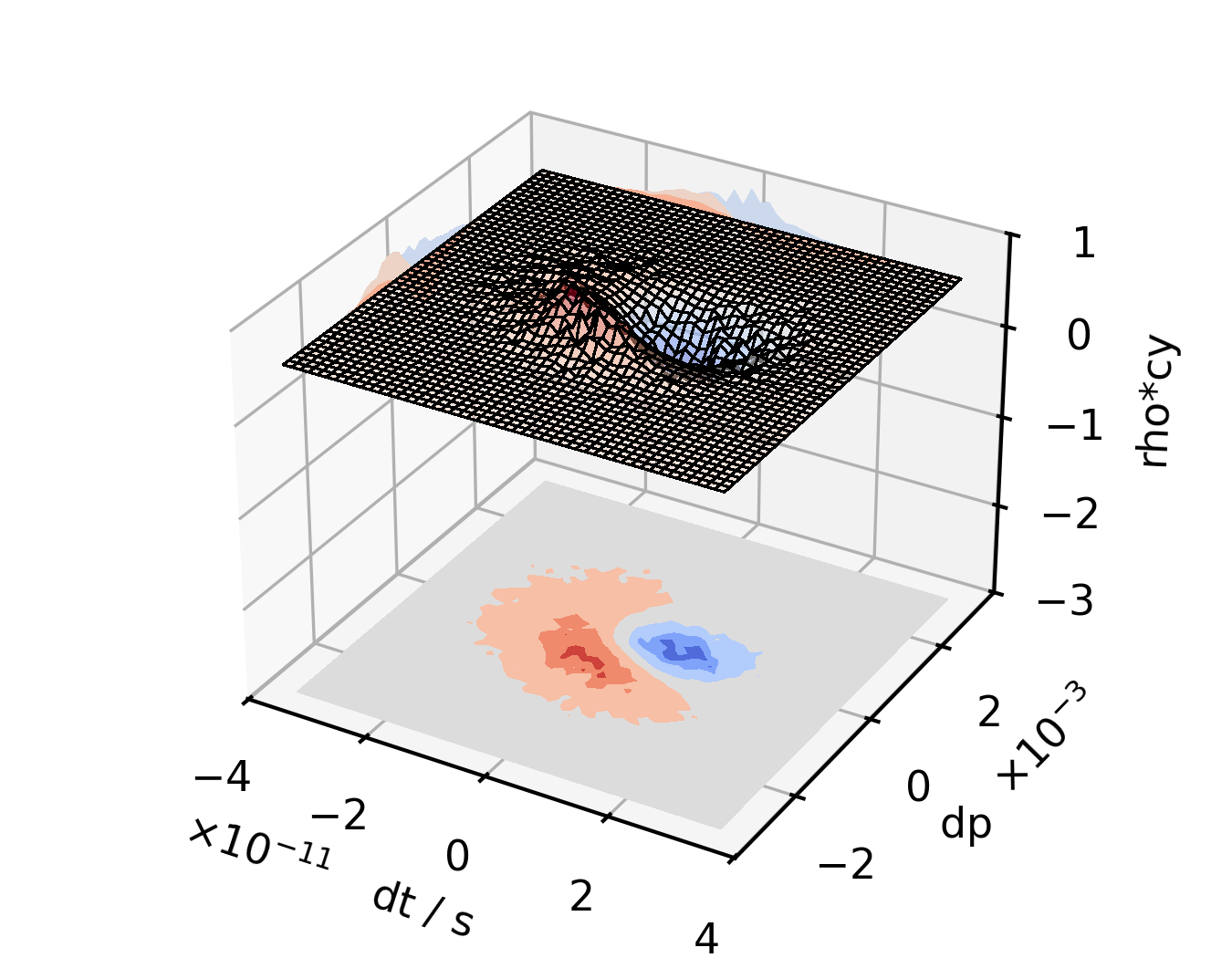}
\includegraphics[width=0.48\textwidth]{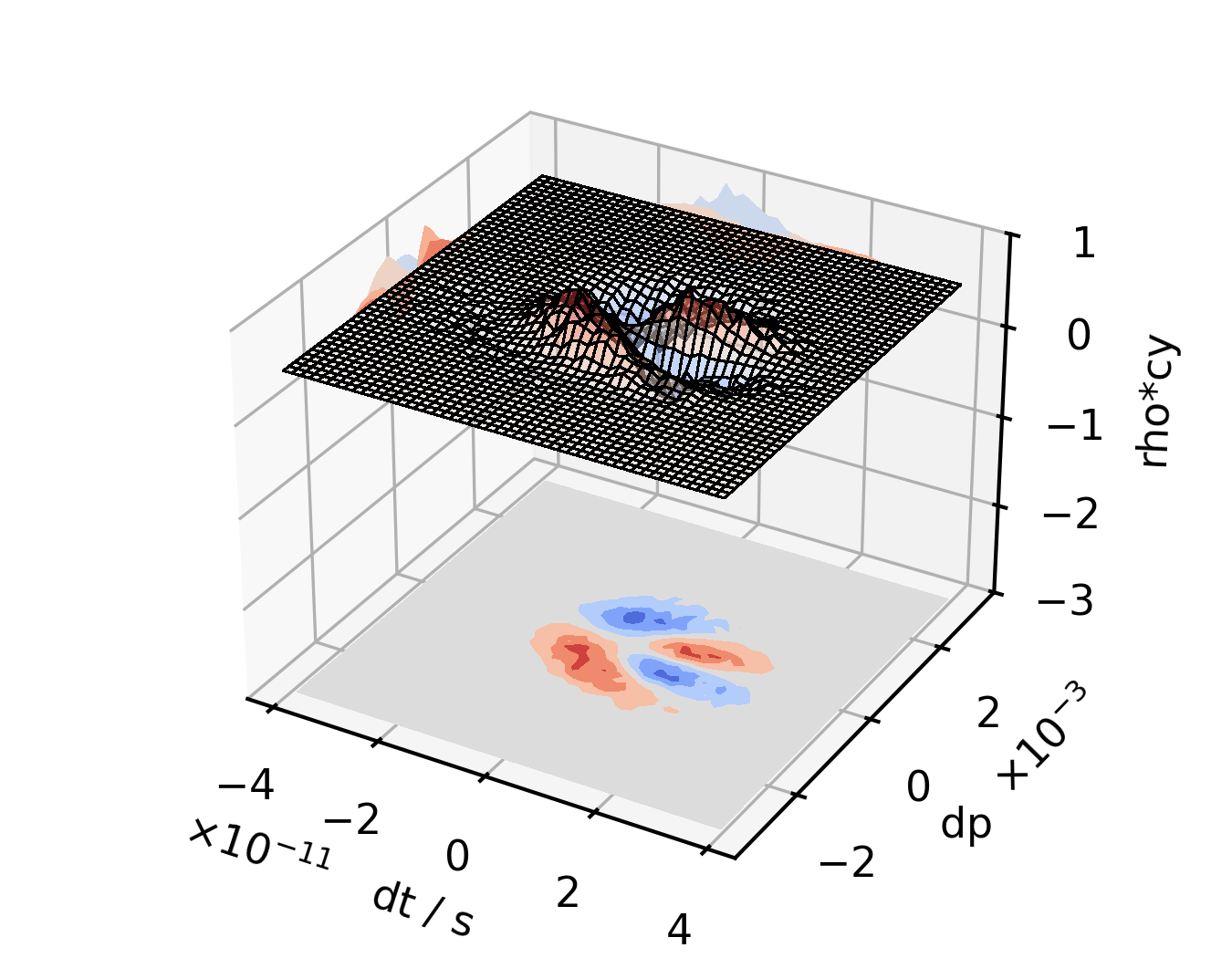}
\includegraphics[width=0.48\textwidth]{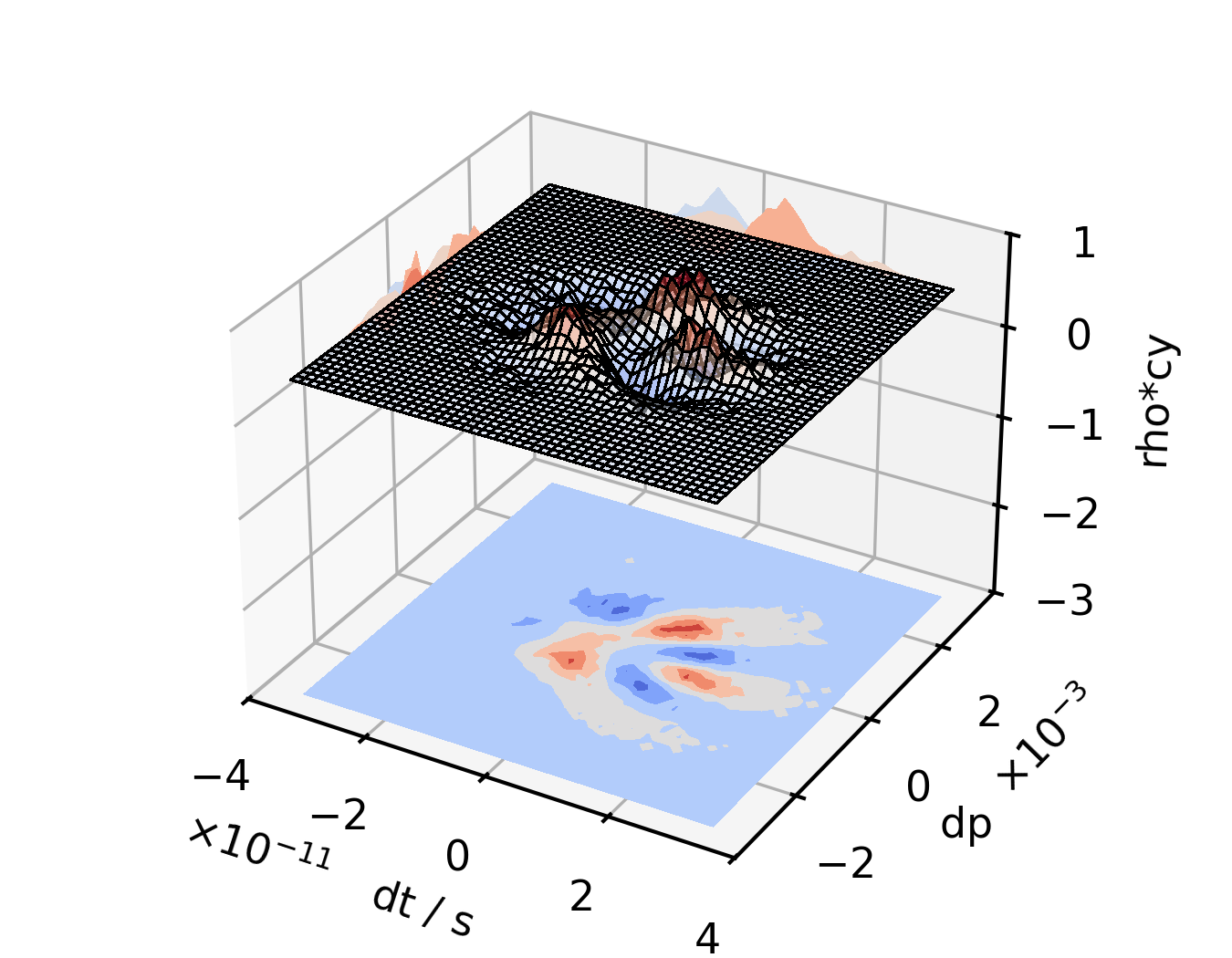}
\caption{Phase-space structures of the dominant modes at different chromaticities. From the top left to the bottom right, the chromaticities are 0, 3, 6, and 9, respectively. The ring is uniformly filled with 80 bunches at a total beam current of 10~mA.}
\label{fig:mode_phase_structure}
\end{figure}

The Vlasov results agree well with the tracking simulations at low chromaticity, while some deviations appear at higher chromaticity. This discrepancy may arise because the tracking data used to extract the growth rates do not remain entirely within the regime of purely exponential growth. Figure~\ref{fig:mode_phase_structure} shows the corresponding phase-space structures at chromaticities of 0, 3, 6, and 9, arranged from the top left to the bottom right. With increasing chromaticity, the azimuthal modes $l=0$, 1, 2, and 3 sequentially become dominant, as expected. Compared with the analytical prediction in Fig.~\ref{fig:ana_mode_phase_structure}, the corresponding Laguerre--Gaussian mode structures are well reproduced in the tracking simulations.

\subsection{Coupled-bunch modes with space charge}\label{sec-IV-B}
In the following study, space-charge effects, synchrotron-radiation damping, and quantum excitation are included in the simulations. As shown in Eq.~(\ref{eq:2.1}), the coupled-bunch instability growth rate scales approximately linearly with the total beam current. According to Tab.~\ref{tab:lattice}, the transverse SR damping rate is approximately $50~\mathrm{s}^{-1}$. To sample unstable modes over a wider chromaticity range, the total beam current is increased to 120~mA by increasing the number of bunches while keeping the single-bunch charge identical to that used in Fig.~\ref{fig:growth_rate_benchmark}.

\begin{figure}[!htp]
\centering
\includegraphics[width=1\textwidth]{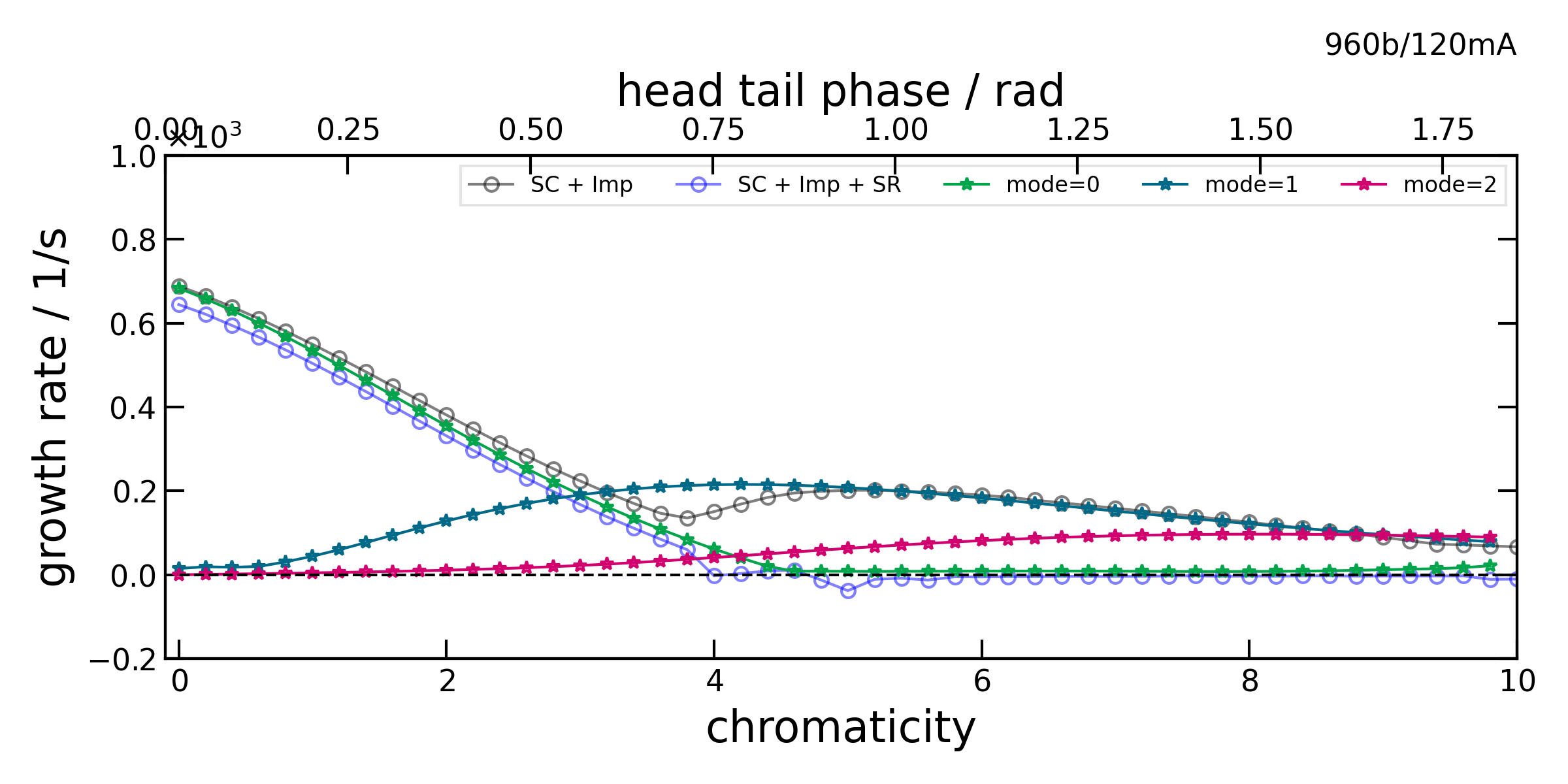}
\caption{Instability growth rates obtained using different analytical and tracking methods. The solid markers represent results from the Vlasov solver, while the open circles represent tracking results with and without radiation damping. The ring is uniformly filled with 960 bunches at a total beam current of 120~mA.}
\label{fig:growth_rate_benchmark_960b_120mA}
\end{figure}

Figure~\ref{fig:growth_rate_benchmark_960b_120mA} shows the corresponding results. The solid markers represent the predictions from the Vlasov solver, with the azimuthal modes $l=0$, 1, and 2 sequentially becoming dominant as the chromaticity increases. Two sets of tracking results, represented by open circles, are shown for comparison, with and without synchrotron-radiation effects. The $l=0$ mode corresponds to rigid-dipole motion cannot be damped by the nonlinear space-charge. When the higher-order modes $l=1, 2$ dominate, the tracking results exhibit a clear bifurcation: the beam is stabilized when both space-charge and SR damping are included, whereas in the absence of SR effects, the growth rates follow the analytical predictions. This behavior reveals a complex interplay between damping and nonlinear space-charge effects. When the transverse beam size is maintained at a smaller value by damping, the resulting higher-density beam core produces a larger nonlinear space-charge tune spread, thereby providing stronger Landau damping. In contrast, if the beam emittance grows significantly from any unstable motion, the reduced beam density weakens the space-charge tune spread and, consequently, the associated Landau damping.

\begin{figure}[!htp]
\centering
\includegraphics[width=0.48\textwidth]{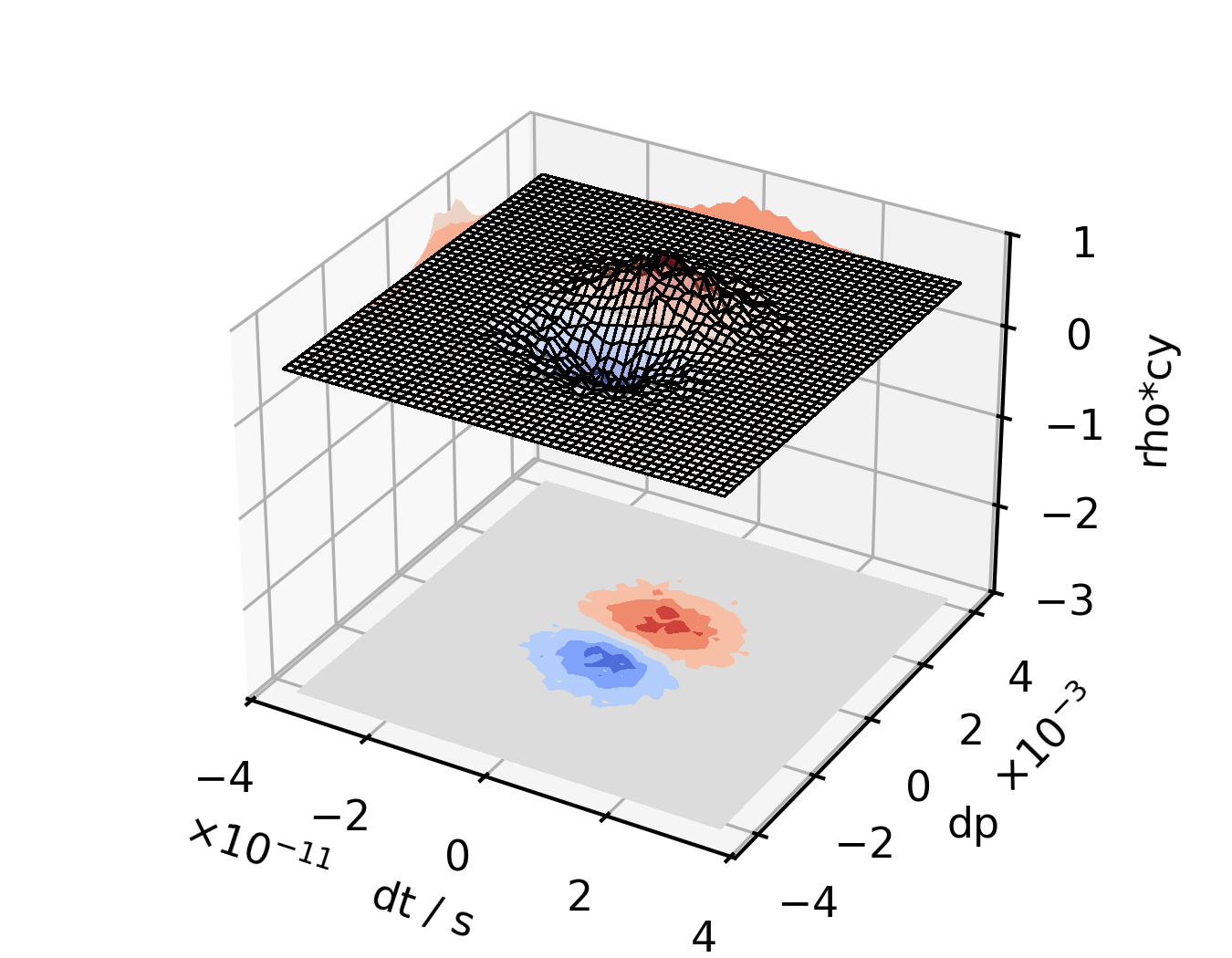}
\includegraphics[width=0.48\textwidth]{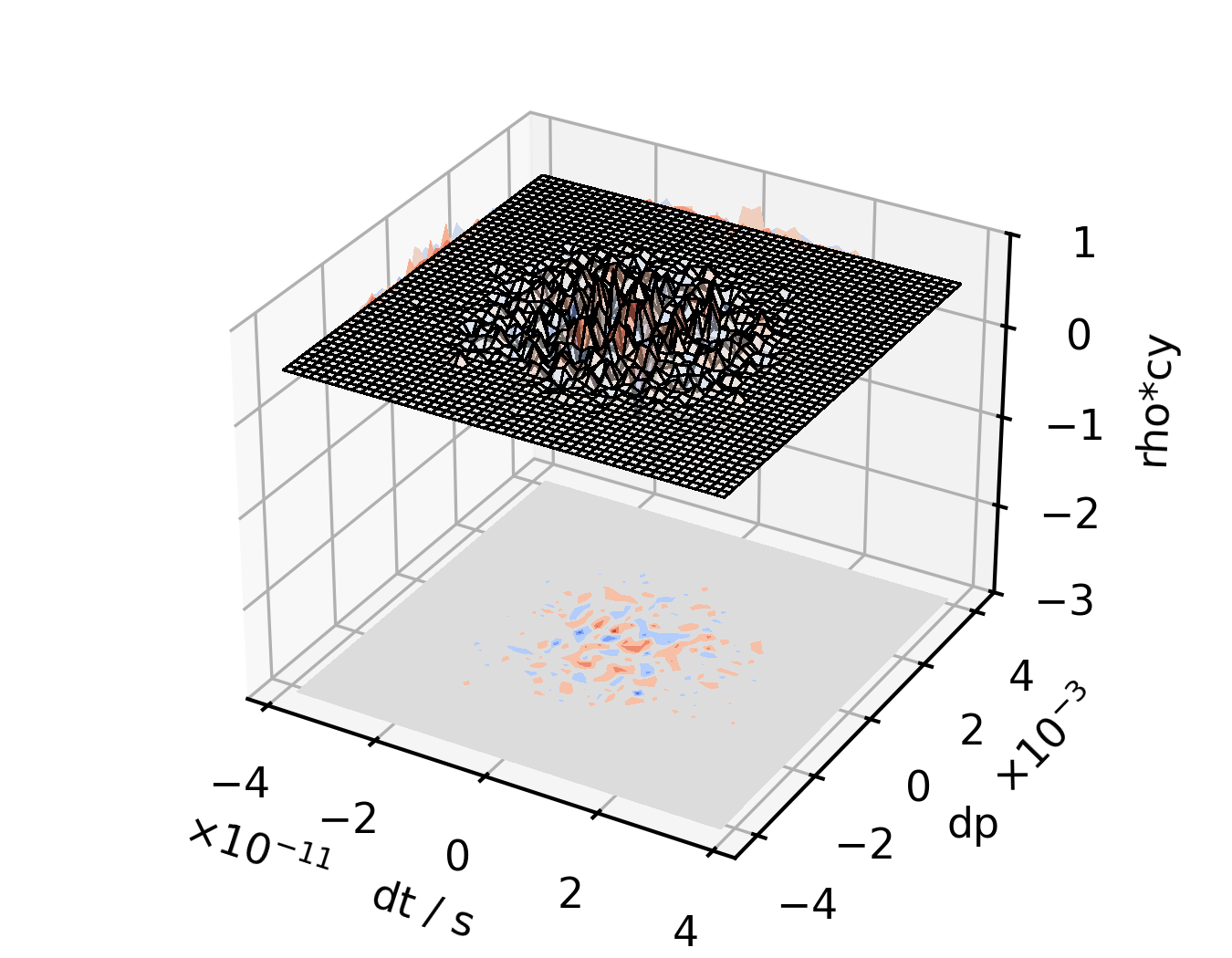}
\caption{Phase-space structures at chromaticity 3 (left) and 6 (right). The ring is uniformly filled with 960 bunches at a total beam current of 120~mA.}
\label{fig:mode_phase_structure_sc}
\end{figure}

As a rough estimate, taking the growth-rate reduction observed around chromaticity 4 as a reference, the effective damping contribution associated with space charge is of the order of $200~\mathrm{s}^{-1}$. For comparison, Fig.~\ref{fig:mode_phase_structure_sc} shows the corresponding phase-space structures with space charge included. The mode structures become progressively smeared out with increasing chromaticity and are almost completely washed out at high chromaticities, providing a clear dynamical signature of mode damping induced by the space charge effect.

\subsection{Tune spread and Landau stability diagram}\label{sec-IV-C}
Equation~(\ref{eq:2-2-4}) provides an analytical estimate of the incoherent tune spread within the frozen bi-Gaussian beam profile. The tune spread can also be extracted directly from particle tracking. One approach is to store the turn-by-turn coordinates of all particles and perform frequency analysis after the beam has reached equilibrium.  Straightforward; however, it requires storing a substantial amount of tracking data. Alternatively, the accumulated phase advance of each particle can be calculated directly in action--angle coordinates during tracking, which can reduce the storage requirement significantly. In both approaches, however, SR damping and quantum excitation introduce stochastic variations in the particle trajectories and can generate considerable numerical noise to reconstruct the tune distribution. The resulting tune footprints should be interpreted with care.

\begin{figure}[!htp]
\centering
\includegraphics[width=1\textwidth]{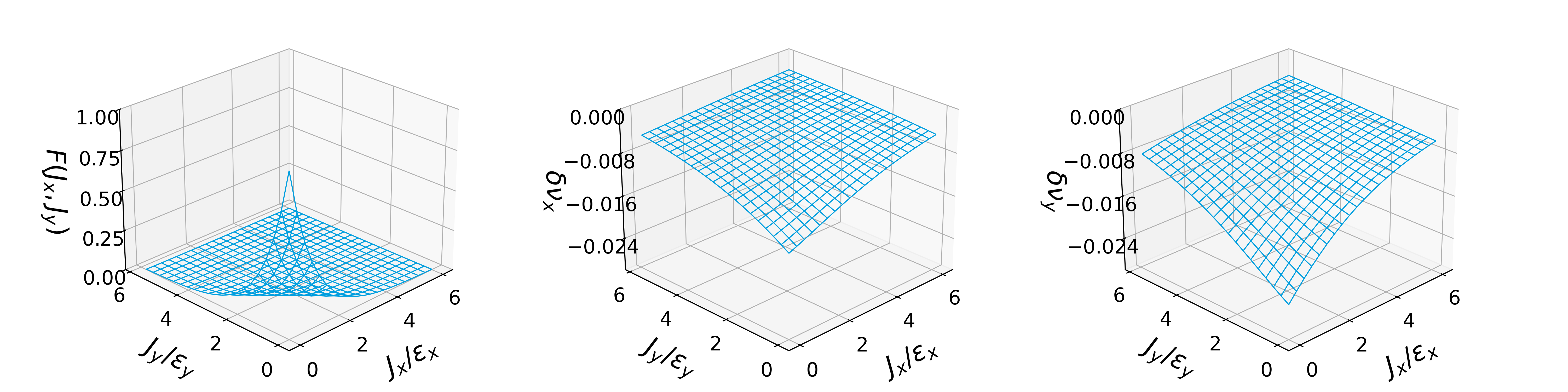}
\caption{Normalized beam distribution and incoherent tune spread in action space $(J_x,J_y)$ for a single-bunch current of 0.125~mA.}
\label{fig:LD_2d_AnaGs_dis_0.125mA}
\end{figure}

\begin{figure}[!htp]
\centering
\includegraphics[width=1\textwidth]{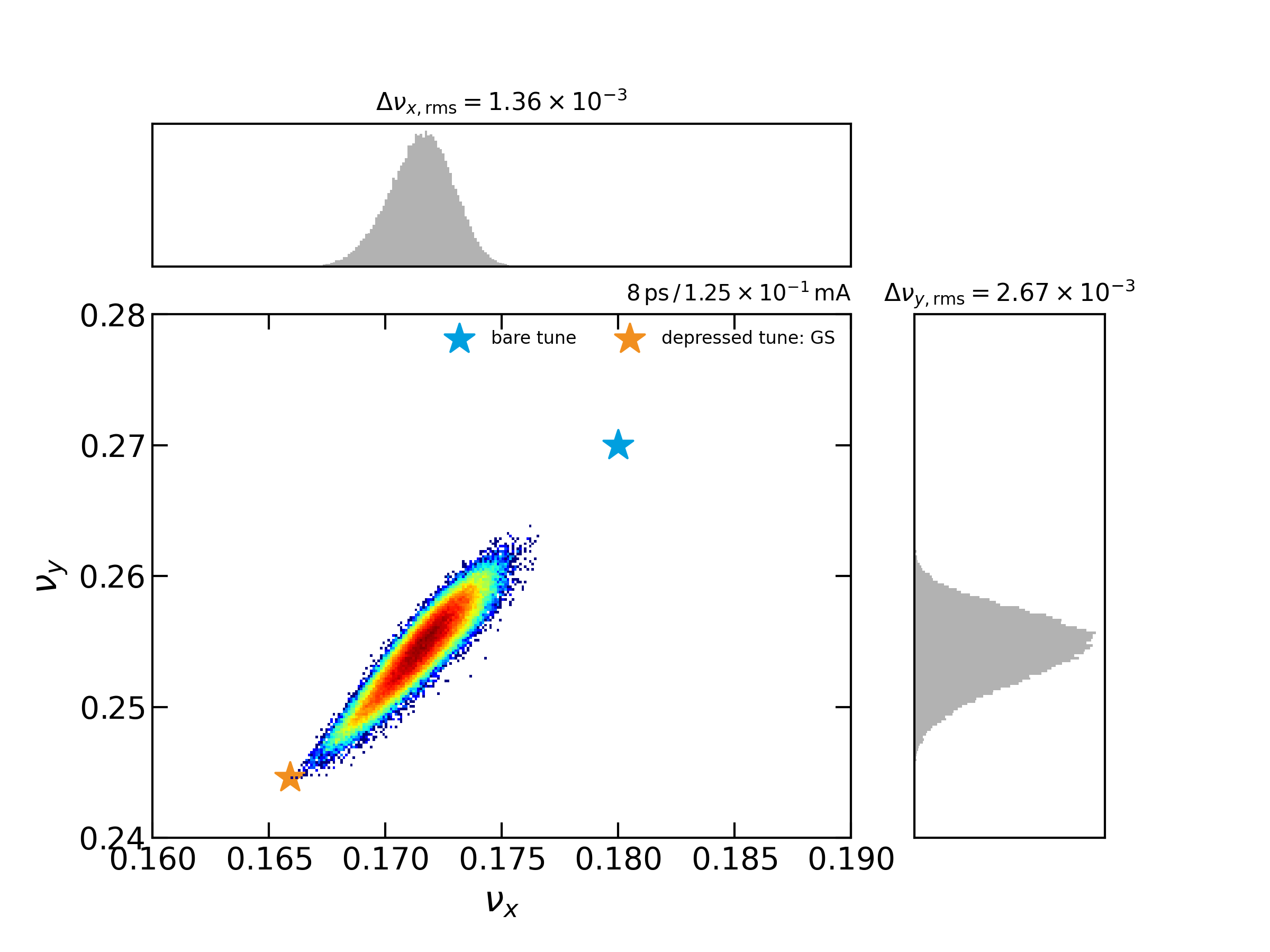}
\caption{Incoherent tune spread due to space charge for a single-bunch current of 0.125~mA. The tune footprint is reconstructed from CETASIM tracking using the accumulated phase-advance method.}
\label{fig:sc_tune_spread_0.125mA}
\end{figure}

Figure~\ref{fig:LD_2d_AnaGs_dis_0.125mA} shows the beam distribution in $(J_x,J_y)$ space and the corresponding incoherent tune spread for a single-bunch current of 0.125~mA, calculated from Eq.~(\ref{eq:2-2-4}). The equilibrium emittances $\epsilon_{u,\mathrm{eq}}$ used in the calculation are obtained from 50,000 turns of particle tracking and are summarized in Tab.~\ref{tab:emit_tune_shift}. For comparison, Fig.~\ref{fig:sc_tune_spread_0.125mA} shows the tune footprint reconstructed from CETASIM tracking using the accumulated phase advance of individual particles. The maximum space-charge tune depression, indicated by the yellow star, agrees well with the analytical prediction~\footnote{Eq.~(\ref{eq:2-2-4}) with $J_x=J_y=0$} in Tab.~\ref{tab:emit_tune_shift}. Only a small fraction of particles extends beyond the 3$J_u$, and consequently, hardly any particles are observed near the bare tune $(0.18,0.27)$. Notably, the rms tune spreads,  $\Delta\nu_u$, in the horizontal and vertical planes are $1.36\times10^{-3}$ and $2.67\times10^{-3}$, respectively, corresponding to characteristic maximum damping rates $2\pi f_0 \Delta \nu_u \sqrt{2\ln 2}$~\cite{chao1993physics} of approximately $1300~\mathrm{s}^{-1}$ and $2600~\mathrm{s}^{-1}$. 

\begin{table}[!htp]
\centering
\caption{Equilibrium emittances and maximum space-charge tune shifts at the beam center.}
\label{tab:emit_tune_shift}
\begin{tabular}{|c|c|c|c|c|} \hline
 & $\epsilon_x$ / pm & $\epsilon_y$ / pm & $\Delta\nu_x$ & $\Delta\nu_y$ \\ \hline
0.125 mA & 18.0 & 9.5 & $1.41\times10^{-2}$ & $2.54\times10^{-2}$ \\ \hline
\end{tabular}
\end{table}

\begin{figure}[!htp]
\centering
\includegraphics[width=1\textwidth]{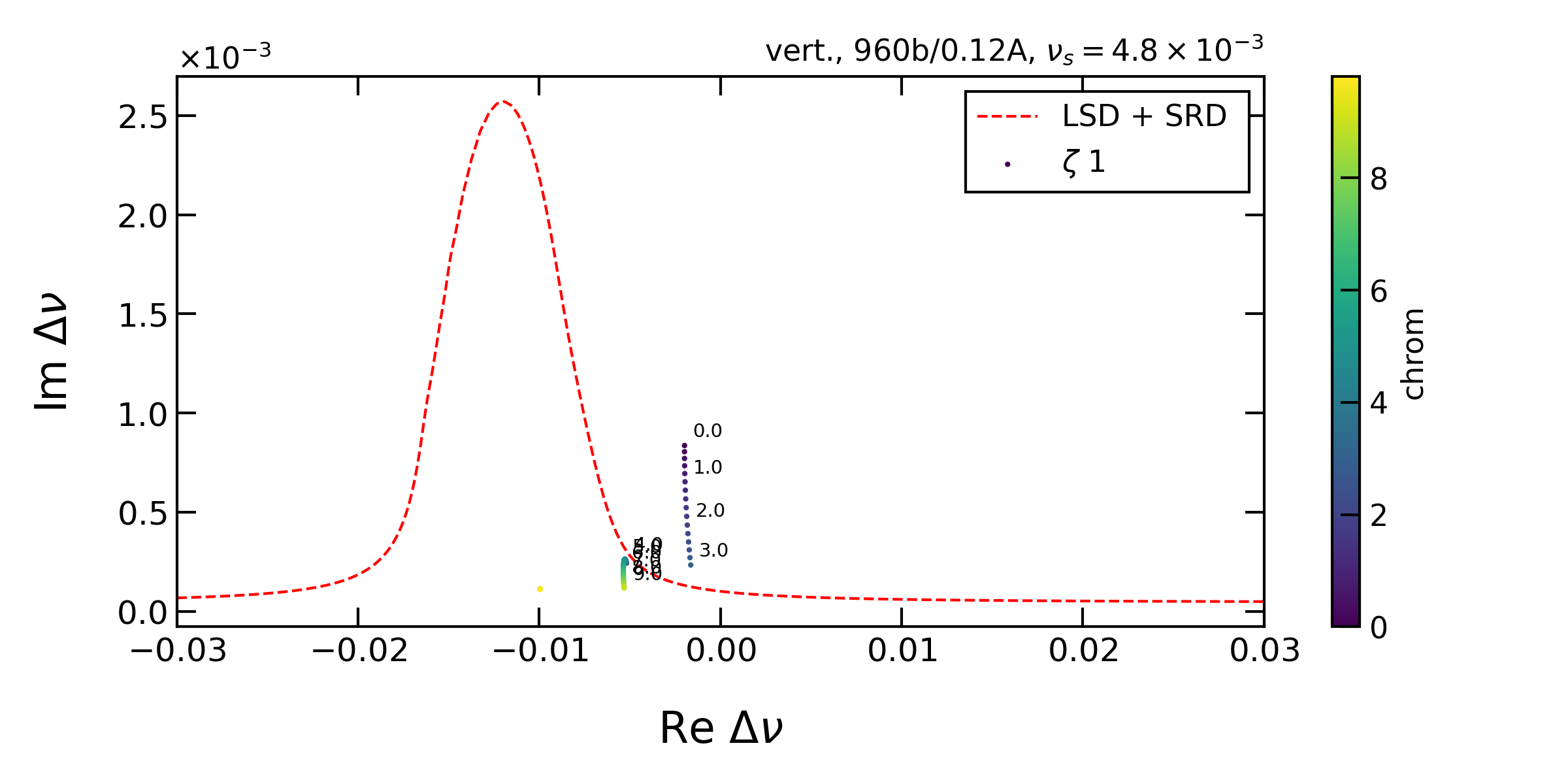}
\caption{Landau stability diagrams in the vertical plane, which are reconstructed from tune spreads obtained by the analytical frozen bi-Gaussian model. $\mathrm{Re}[\Delta\nu]=0$ corresponds to the bare tune.}
\label{fig:landau_damping_contour}
\end{figure}

The red curve in Fig.~\ref{fig:landau_damping_contour} represents the Landau stability diagram obtained from the tune spread predicted by the frozen bi-Gaussian model. The real and imaginary parts of $\Delta\nu$ represent the coherent mode frequency shift and growth rate, respectively, with $\operatorname{Re}(\Delta\nu)=0$ corresponding to the bare tune. The colored dots represent the dominant coupled-bunch modes as the chromaticity is increased. The modes associated with azimuthal index $l$ are clustered around $\operatorname{Re}(\Delta\nu)=-l\nu_s$. These points contain the same growth-rate information as that shown in Fig.~\ref{fig:growth_rate_benchmark_960b_120mA}, while additionally providing the corresponding coherent frequency shifts. It can be clearly seen that the higher-order modes ($l\geq1$) lie within the stable region, explaining the behavior observed in Fig.~\ref{fig:growth_rate_benchmark_960b_120mA}. The maximum damping rate provided by the Landau stability diagram is approximately $\operatorname{Im}(\Delta\nu)=2.6\times10^{-3}$, which is in reasonable agreement with the estimate based on the rms incoherent tune spread.

The data set shown in Fig.~\ref{fig:sc_tune_spread_0.125mA} can be applied to Eq.~(\ref{eq:2-3-1}) to construct the corresponding Landau stability diagram as well. A sufficiently long tracking is required to determine the individual particle tunes with adequate accuracy, whereas more noise would be introduced from the SR damping and quantum excitation for tune spread reconstruction. Consequently, obtaining a converged Landau stability diagram directly from particle-tracking data is challenging in an electron storage ring. As a cross-check, the same reconstruction method was also tested for a lattice with amplitude-dependent tune shifts generated by octupoles, with radiation damping and quantum excitation switched off during particle tracking. In this case, the corresponding Landau stability diagram can be reliably reconstructed and agrees well with the analytical prediction. This comparison indicates that direct reconstruction of the incoherent tune spread from particle tracking is more straightforward in systems where stochastic radiation effects are negligible, as is typically the case for hadron machines.

\section{Discussion and conclusion}\label{sec-V}

Although space-charge effects are strongly suppressed at high beam energies, the extremely small transverse emittances of diffraction-limited storage rings result in sufficiently high charge densities that the associated nonlinear tune shifts and tune spreads can no longer be neglected. In this work, we investigated the impact of direct space charge on the transverse coupled-bunch instability in PETRA-IV. To simplify the analysis and facilitate a direct comparison with analytical predictions, a simplified resistive-wall (RW) impedance model was adopted, while bunch lengthening from the harmonic cavity and the longitudinal broadband impedance were not included.

To model space charge efficiently, a 2.5D solver based on the Bassetti--Erskine formulation was benchmarked between \textsc{Elegant} and CETASim. The two codes show good agreement in the overall evolution of the beam emittance. A simplified one-turn-map model was subsequently introduced for multi-bunch simulations, allowing the nonlinear space-charge contribution to be isolated from other lattice nonlinearities while substantially reducing the computational cost. Element-by-element tracking, however, remains necessary when emittance growth or detailed incoherent resonance structures are considered as figures of merit.

When space charge, synchrotron-radiation damping, and quantum excitation are included simultaneously, a strong interplay between radiation damping and space-charge-induced Landau damping is observed. Radiation damping maintains a dense beam core and, consequently, a large nonlinear space-charge tune spread, thereby enhancing the Landau damping of higher-order coupled-bunch modes. In contrast, in the absence of radiation damping, the increase in beam size reduces the charge density, weakens the space-charge tune spread, and consequently diminishes the associated Landau damping. The incoherent tune spread was evaluated both analytically using a frozen bi-Gaussian model and numerically from particle tracking. The stability diagram predicted by the analytical model shows the stabilization of higher-order coupled-bunch modes, in agreement with the tracking simulations. The maximum tune depression extracted from tracking also agrees well with the analytical prediction. However, reconstructing the full incoherent tune footprint from long-term particle tracking is challenging in an electron storage ring because of synchrotron-radiation damping and quantum excitation. A dedicated cross-check using octupole-induced amplitude-dependent tune shifts, with radiation damping and quantum excitation switched off, confirms the validity of the tracking-based reconstruction method in the absence of these stochastic radiation effects.

Overall, with PETRA-IV as an example, the results demonstrate that the nonlinear tune spread induced by direct space charge can provide significant Landau damping of transverse coupled-bunch modes in a diffraction-limited electron storage ring. The approach and conclusions of the present study can also be extended to other diffraction-limited storage rings where the combination of small transverse emittance and high bunch charge gives rise to appreciable direct space-charge effects.

\section{ACKNOWLEDGMENTS}\label{sec-VI}
The author would like to take this opportunity to warmly thank the PETRA~IV Beam Physics Group for its support and many fruitful discussions. In particular, the author would like to thank Yong-Chul Chae for many helpful discussions and numerous valuable comments and suggestions. The author is also grateful to Sergey Antipov for valuable discussions on the Landau stability diagram and for carefully proofreading the manuscript. This research was supported by the Maxwell computational resources operated at Deutsches Elektronen-Synchrotron DESY, Hamburg, Germany.

\section{Appendix}
The potential from the space-charge can be written in the format~\cite{takayama1982new,ng2006physics,PhysRevAccelBeams.25.121001} 
\begin{equation}\label{eq:A-1}
U(x,y,z;s) = \frac{N_p e} {4\pi\varepsilon_0\sqrt{\pi}}
\int_{0}^{\infty}\frac{\exp\left(-\frac{x^2}{2\sigma_x^2+q}-\frac{y^2}{2\sigma_y^2+q}
-\frac{z^2}{2\sigma_z^2+q}\right)}{
\sqrt{\left(2\sigma_x^2+q\right)\left(2\sigma_y^2+q\right)\left(2\sigma_z^2+q\right)
}}\,\mathrm{d}q .
\end{equation}
where $\sigma_{x,y,z}$ are the rms beam size, $N_p$ is the electron number. In the long bunch region \cite{PhysRevAccelBeams.27.094201,4440457}, the integration along the longitudinal can be take out. If we further take $\lambda(z) = N_p \exp(-z^2/2\sigma_z^2) / \sqrt{2 \pi} \sigma_z $ as the line density, the transverse space-charge potential can be simplified to 
\begin{equation}\label{eq:A-2}
U(x,y;s) = \frac{\lambda e} {4\pi\varepsilon_0}
\int_{0}^{\infty}\frac{\exp\left(-\frac{x^2}{2\sigma_x^2+q}-\frac{y^2}{2\sigma_y^2+q}\right)}{\sqrt{\left(2\sigma_x^2+q\right)\left(2\sigma_y^2+q\right)}}\,\mathrm{d}q,
\end{equation}
which is the same as the Bassetti-Eskin formula~\cite{Bassetti}. Hereafter, treating the space-charge as a perturbation, the Hamiltonian of the particle can be expressed as 
\begin{equation}\label{eq:A-3}
H(x,x',y,y';s) = \frac{1}{2}(x'^2 + K_x(s) x^2 + y'^2 + K_y(s) y^2) + \frac{U(x,y;s)}{m_ec^2 \beta^2 \gamma^3} 
\end{equation}

\bibliographystyle{ieeetr}
\bibliography{note}
\end{document}